
%
%
\input harvmac %
%
%
%
%
%
%
%
%
%
\newif\ifdraft

\noblackbox
\catcode`\@=11
\newif\iffrontpage
%
\ifx\answ\bigans
\def\titleft{\titsm}
\magnification=1200\baselineskip=15pt plus 2pt minus 1pt
%
\voffset=0.35truein
\advance\hoffset by-0.075truein
\hsize=6.15truein\vsize=600.truept\hsbody=\hsize\hstitle=\hsize
\else\let\lr=L
\def\titleft{\titla}
\magnification=1000\baselineskip=14pt plus 2pt minus 1pt
%
\vsize=6.5truein
\hstitle=8truein\hsbody=4.75truein
\fullhsize=10truein\hsize=\hsbody
\fi
\parskip=4pt plus 10pt minus 4pt

\font\titla=cmr10 scaled\magstep3
\font\tenmss=cmss10
\font\absmss=cmss10 scaled\magstep1
\newfam\mssfam
\font\footrm=cmr8  \font\footrms=cmr5
\font\footrmss=cmr5   \font\footi=cmmi8
\font\footis=cmmi5   \font\footiss=cmmi5
\font\footsy=cmsy8   \font\footsys=cmsy5
\font\footsyss=cmsy5   \font\footbf=cmbx8
\font\footmss=cmss8
\def\footfont{\def\rm{\fam0\footrm}
\textfont0=\footrm \scriptfont0=\footrms
\scriptscriptfont0=\footrmss
\textfont1=\footi \scriptfont1=\footis
\scriptscriptfont1=\footiss
\textfont2=\footsy \scriptfont2=\footsys
\scriptscriptfont2=\footsyss
\textfont\itfam=\footi \def\it{\fam\itfam\footi}
\textfont\mssfam=\footmss \def\mss{\fam\mssfam\footmss}
\textfont\bffam=\footbf \def\bf{\fam\bffam\footbf} \rm}
\def\tenpoint{\def\rm{\fam0\tenrm}
\textfont0=\tenrm \scriptfont0=\sevenrm
\scriptscriptfont0=\fiverm
\textfont1=\teni  \scriptfont1=\seveni
\scriptscriptfont1=\fivei
\textfont2=\tensy \scriptfont2=\sevensy
\scriptscriptfont2=\fivesy
\textfont\itfam=\tenit \def\it{\fam\itfam\tenit}
\textfont\mssfam=\tenmss \def\mss{\fam\mssfam\tenmss}
\textfont\bffam=\tenbf \def\bf{\fam\bffam\tenbf} \rm}
\ifx\answ\bigans\def\abstractfont{\tenpoint}\else
\def\abstractfont{\def\rm{\fam0\absrm}
\textfont0=\absrm \scriptfont0=\absrms
\scriptscriptfont0=\absrmss
\textfont1=\absi \scriptfont1=\absis
\scriptscriptfont1=\absiss
\textfont2=\abssy \scriptfont2=\abssys
\scriptscriptfont2=\abssyss
\textfont\itfam=\bigit \def\it{\fam\itfam\bigit}
\textfont\mssfam=\absmss \def\mss{\fam\mssfam\absmss}
\textfont\bffam=\absbf \def\bf{\fam\bffam\absbf}\rm}\fi
%
\def\f@@t{\baselineskip10pt\lineskip0pt\lineskiplimit0pt
\bgroup\aftergroup\@foot\let\next}
\setbox\strutbox=\hbox{\vrule height 8.pt depth 3.5pt width\z@}
\def\vfootnote#1{\insert\footins\bgroup
\baselineskip10pt\footfont
\interlinepenalty=\interfootnotelinepenalty
\floatingpenalty=20000
\splittopskip=\ht\strutbox \boxmaxdepth=\dp\strutbox
\leftskip=24pt \rightskip=\z@skip
\parindent=12pt \parfillskip=0pt plus 1fil
\spaceskip=\z@skip \xspaceskip=\z@skip
\Textindent{$#1$}\footstrut\futurelet\next\fo@t}
\def\Textindent#1{\noindent\llap{#1\enspace}\ignorespaces}
\def\footnote#1{\attach{#1}\vfootnote{#1}}%

\def\foot{\attach\footsymbolgen\vfootnote{\footsymbol}}
\let\footsymbol=\star
\newcount\lastf@@t           \lastf@@t=-1
\newcount\footsymbolcount    \footsymbolcount=0
\def\footsymbolgen{\relax\footsym
\global\lastf@@t=\pageno\footsymbol}
\def\footsym{\ifnum\footsymbolcount<0
\global\footsymbolcount=0\fi
{\iffrontpage \else \advance\lastf@@t by 1 \fi
\ifnum\lastf@@t<\pageno \global\footsymbolcount=0
\else \global\advance\footsymbolcount by 1 \fi }
\ifcase\footsymbolcount \fd@f\star\or
\fd@f\dagger\or \fd@f\ast\or
\fd@f\ddagger\or \fd@f\natural\or
\fd@f\diamond\or \fd@f\bullet\or
\fd@f\nabla\else \fd@f\dagger
\global\footsymbolcount=0 \fi }
\def\fd@f#1{\xdef\footsymbol{#1}}
\def\space@ver#1{\let\@sf=\empty \ifmmode #1\else \ifhmode
\edef\@sf{\spacefactor=\the\spacefactor}
\unskip${}#1$\relax\fi\fi}
\def\attach#1{\space@ver{\strut^{\mkern 2mu #1}}\@sf}
%
\newif\ifnref
\def\rrr#1#2{\relax\ifnref\nref#1{#2}\else\ref#1{#2}\fi}
\def\ldf#1#2{\begingroup\obeylines
\gdef#1{\rrr{#1}{#2}}\endgroup\unskip}
\def\nrf#1{\nreftrue{#1}\nreffalse}
\def\doubref#1#2{\refs{{#1},{#2}}}

\nreffalse
\def\refout{\listrefs}
%
\def\eqn#1{\xdef #1{(\secsym\the\meqno)}
\writedef{#1\leftbracket#1}%
\global\advance\meqno by1\eqno#1\eqlabeL#1}
\def\eqnalign#1{\xdef #1{(\secsym\the\meqno)}
\writedef{#1\leftbracket#1}%
\global\advance\meqno by1#1\eqlabeL{#1}}
%
\def\chap#1{\newsec{#1}}
\def\chapter#1{\chap{#1}}
\def\sect#1{\subsec{{ #1}}}
\def\section#1{\sect{#1}}
\def\\{\ifnum\lastpenalty=-10000\relax
\else\hfil\penalty-10000\fi\ignorespaces}
\def\note#1{\leavevmode%
\edef\@@marginsf{\spacefactor=\the\spacefactor\relax}%
\ifdraft\strut\vadjust{%
\hbox to0pt{\hskip\hsize%
\ifx\answ\bigans\hskip.1in\else\hskip-.1in\fi%
\vbox to0pt{\vskip-\dp
\strutbox\sevenbf\baselineskip=8pt plus 1pt minus 1pt%
\ifx\answ\bigans\hsize=.7in\else\hsize=.35in\fi%
\tolerance=5000 \hbadness=5000%
\leftskip=0pt \rightskip=0pt \everypar={}%
\raggedright\parskip=0pt \parindent=0pt%
\vskip-\ht\strutbox\noindent\strut#1\par%
\vss}\hss}}\fi\@@marginsf\kern-.01cm}
\def\titlepage{%
\frontpagetrue\nopagenumbers\abstractfont%
\hsize=\hstitle\rightline{\vbox{\baselineskip=10pt%
{\abstractfont\pubnum}}}\pageno=0}
\frontpagefalse
\def\pubnum{}
\def\pdate{\number\month/\number\yearltd}
\def\makefootline{\iffrontpage\vskip .27truein
\line{\the\footline}
\vskip -.1truein\leftline{\vbox{\baselineskip=10pt%
{\abstractfont\pdate}}}
\else\vskip.5cm\line{\hss \tenrm $-$ \folio\ $-$ \hss}\fi}
\def\title#1{\vskip .7truecm\titlestyle{\titleft #1}}
\def\titlestyle#1{\par\begingroup \interlinepenalty=9999
\leftskip=0.02\hsize plus 0.23\hsize minus 0.02\hsize
\rightskip=\leftskip \parfillskip=0pt
\hyphenpenalty=9000 \exhyphenpenalty=9000
\tolerance=9999 \pretolerance=9000
\spaceskip=0.333em \xspaceskip=0.5em
\noindent #1\par\endgroup }
\def\autskip{\ifx\answ\bigans\vskip.5truecm\else\vskip.1cm\fi}
\def\author#1{\vskip .7in \centerline{#1}}

\def\address#1{\ifx\answ\bigans\vskip.2truecm
\else\vskip.1cm\fi{\it \centerline{#1}}}
\def\abstract#1{
\vskip .5in\vfil\centerline
{\bf Abstract}\penalty1000
{{\smallskip\ifx\answ\bigans\leftskip 2pc \rightskip 2pc
\else\leftskip 5pc \rightskip 5pc\fi
\noindent\abstractfont \baselineskip=12pt
{#1} \smallskip}}
\penalty-1000}
\def\endpage{\tenpoint\supereject\global\hsize=\hsbody%
\frontpagefalse\footline={\hss\tenrm\folio\hss}}
\def\ack{\goodbreak\vskip2.cm\centerline{{\bf Acknowledgements}}}
\def\CERN{\address{CERN, CH--1211 Geneva 23, Switzerland}}
\def\bfone{\relax{\rm 1\kern-.35em 1}}
\def\inbar{\vrule height1.5ex width.4pt depth0pt}
\def\IC{\relax\,\hbox{$\inbar\kern-.3em{\mss C}$}}
\def\ID{\relax{\rm I\kern-.18em D}}
\def\IF{\relax{\rm I\kern-.18em F}}
\def\IH{\relax{\rm I\kern-.18em H}}
\def\II{\relax{\rm I\kern-.17em I}}
\def\IN{\relax{\rm I\kern-.18em N}}
\def\IP{\relax{\rm I\kern-.18em P}}
\def\IQ{\relax\,\hbox{$\inbar\kern-.3em{\rm Q}$}}
\def\IR{\relax{\rm I\kern-.18em R}}
\font\cmss=cmss10 \font\cmsss=cmss10 at 7pt
\def\ZZ{\relax\ifmmode\mathchoice
{\hbox{\cmss Z\kern-.4em Z}}{\hbox{\cmss Z\kern-.4em Z}}
{\lower.9pt\hbox{\cmsss Z\kern-.4em Z}}
{\lower1.2pt\hbox{\cmsss Z\kern-.4em Z}}\else{\cmss Z\kern-.4em
Z}\fi}
\def\a{\alpha} \def\b{\beta} \def\d{\delta}
 \def\c{\gamma}

\def\cA{{\cal A}}

 \def\cK{{\cal K}}
 \def\cM{{\cal M}}

\def\nup#1({Nucl.\ Phys.\ $\us {B#1}$\ (}
\def\plt#1({Phys.\ Lett.\ $\us  {#1}$\ (}
\def\cmp#1({Comm.\ Math.\ Phys.\ $\us  {#1}$\ (}
\def\prp#1({Phys.\ Rep.\ $\us  {#1}$\ (}
\def\prl#1({Phys.\ Rev.\ Lett.\ $\us  {#1}$\ (}
\def\prv#1({Phys.\ Rev.\ $\us  {#1}$\ (}
\def\mpl#1({Mod.\ Phys.\ \Let.\ $\us  {#1}$\ (}
\def\tit#1|{{\it #1},\ }
%

%

\def\tilde{\widetilde}
\def\bar{\overline}
\def\us#1{\underline{#1}}

\def\hat{\widehat}

\def\Coe#1.#2.{{#1\over #2}}
\def\coeff#1#2{\relax{\textstyle {#1 \over #2}}\displaystyle}
\def\coe#1.#2.{\relax{\textstyle {#1 \over #2}}\displaystyle}

\def\shalf{\relax{\textstyle {1 \over 2}}\displaystyle}

\def\to{\rightarrow}
\def\notin{\hbox{{$\in$}\kern-.51em\hbox{/}}}

\def\del{\partial}

\catcode`\@=12

\ldf\who{Who should be referenced here??xx}
%
%
\ldf\ADS{I.~Affleck, M.~Dine and N.~Seiberg, Phys. Rev.
  Lett.  51 (1984) 1026, \nup241 (1984) 493, \nup256 (1985) 557.}
\ldf\ANT{
I.~Antoniadis, K.~Narain and T.~Taylor, \plt B267 (1991) 37.}
\ldf\BFOFW{J.\ Balog, L.\ Feher, L.\ O'Raifeartaigh, P.\ Forga\'cs
and A.\ Wipf, \plt B244 (1990) 435; Ann.\ Phys.\ $\us{203}$ (1990) 194.}
\ldf\BDFS{T.\ Banks,
L.\ Dixon, D.\ Friedan and S.\ Shenker, \nup299 (1988)
 613.}
\ldf\BGa{P. Bin\'etruy and M. K. Gaillard, \nup\
B358 (1991) 121.}
\ldf\BGb{P. Bin\'etruy and M. K. Gaillard, \plt   B253
(1990) 119.}
\ldf\BGG{P. Binetruy, G. Girardi and R. Grimm,
LAPP preprint LAPP-TH-337-91.}
\ldf\BV{B. Blok and  A.\ Varchenko, Int.J.Mod.Phys.A7, (1992) 1467.}
\ldf\CF  {A.\ Cadavid and S.\ Ferrara, \plt B267 (1991) 193.}
\ldf\Candelas{P.\ Candelas, \nup 298 (1988) 458.}
\ldf\CD{P.\ Candelas and X.C.\ de la Ossa, \nup355 (1991) 455.}
\ldf\CDGP  {P.\ Candelas, X.C.\ de la Ossa, P.S.\ Green and
L.\ Parkes, \plt B258 (1991) 118; \nup359 (1991) 21.}
\ldf\CHSW{P.~Candelas, G.~Horowitz,
  A.~Strominger and E.~Witten, \nup258 (1985) 46.}
\ldf\CLO{G. Cardoso Lopes and B. Ovrut, UPR-preprint 0464T }
\ldf\CLMR{
J.~A.~Casas, Z.~Lalak, C.~Mu\~noz and G.G.~Ross,
\nup347 (1990) 243.}
\ldf\Cecotti{S.\ Cecotti,
 \nup 355 (1991) 755, Int.\ J.\ Mod.\ Phys.\ A6 (1991) 1749.}
\ldf\CFG{
S.\ Cecotti, S.\ Ferrara and L.\ Girardello,
Int.\ Mod.\ J.\ Phys.\ A4 (1989) 2475, \plt B213 (1988) 443.}
\ldf\CFV{S. Cecotti, S. Ferrara and M. Villasante, Int. J. Mod.
Phys.  A2 (1987) 1839.}
\ldf\CV  {S.\ Cecotti and C.\ Vafa, \nup367 (1991) 359.}
\ldf\CDFLL{A.~Ceresole, R.~D'Auria, S.~Ferrara, W.~Lerche and
J.~Louis, {\it \pf\ equations and special geometry},
CERN preprint CERN-TH.6441/92.}
\ldf\Cvetic{For a review see M.~Cveti\v c, {\it in} Proceedings of
Trieste Summer School in High Energy Physics and Cosmology,
Trieste, Italy, 1988 and 1987, and references therein.}
\ldf\DFKZ{J. P. Derendinger, S. Ferrara, C. Kounnas and F. Zwirner,
Cern preprint TH.6004/91}
\ldf\DIN{J.P.~Derendinger, L.E.~Ib\'a\~nez and
  H.P.~Nilles, \plt  155B (1985) 65.}
\ldf\dspert{M.~Dine and
  N.~Seiberg, Phys. Rev. Lett.  57 (1986) 2625.}
\ldf\dines{M.~Dine and N.~Seiberg,
  in {\it Unified String Theories},
  eds. M.~Green and D.~Gross (World Scientific, 1986),
  \plt  162B (1985) 299.}
\ldf\DVV {R.\ Dijkgraaf, E. Verlinde and H. Verlinde,
\nup352(1991) 59.}
\ldf\DIZ  {P.\ Di Francesco, C.\ Itzykson and J.-B.\ Zuber,
\cmp140 (1991) 543.}
\ldf\DRSW{M.~Dine, R.~Rohm, N.~Seiberg and
  E.~Witten, \plt  156B (1985) 55.}
\ldf\DS{M.~Dine and N.~Seiberg, Phys. Rev. Lett.
    55 (1985) 366.}
\ldf\dixontrieste{For a review see
L. Dixon, {\sl in} Proc. of the 1987 ICTP Summer
Workshop in High Energy Physics, Trieste, Italy, ed.~ G.~Furlan,
J.~C.~Pati, D.~W.~Sciama, E.~Sezgin and Q.~Shafi
and references therein.}
\ldf\dixonberkeley{L. Dixon, talk presented at the M.S.R.I. meeting
on
mirror symmetries, Berkeley, May 1991.}
\ldf\dixondpf{L. Dixon, talk presented at the A.P.S. D.P.F.
  Meeting, Houston, 1990, SLAC-PUB 5229.}
\ldf\DFMS{
  S.~Hamidi and C.~Vafa, \nup279 (1987) 465;\brk
  L.~Dixon, D.~Friedan, E.~Martinec and S.~Shenker,
  \nup282 (1987) 13.}
\ldf\DKLa{L.J.\ Dixon, V.S.\ Kaplunovsky and J.\ Louis,
  \nup329 (1990) 27.}
\ldf\DKLb{L.~Dixon, V.~Kaplunovsky and J.~Louis, \nup355 (1991) 649.}
\ldf\DKLP{
L.~Dixon, V.~Kaplunovsky, J.~Louis and M.~Peskin, unpublished.}
\ldf\DS{Drinfel'd and V.~G.~Sokolov,
 Jour.~Sov.~Math. {\bf 30} (1985) 1975.}
\ldf\dubrovin{B. Dubrovin, {\it Geometry and integrability of
   topological--antitopological fusion}, preprint
   INFN-8-92-DSF; {\it Integrable systems in topological
   field theory}, preprint INFN-AE-92-01 (Jan 1992).}
\ldf\FFS{S.\ Ferrara,
P.\  Fr\`e and P.\ Soriani, {\it On the moduli space
of the $T^6/Z_3$ orbifold and its modular group}, preprint CERN-TH
6364/92,
SISSA 5/92/EP.}
\ldf\FGN{S. Ferrara, L. Girardello and H. P. Nilles,
\plt  125B (1983) 457}
\ldf\FGPS{S. Ferrara, L.Girardello, O. Piguet and R. Stora,
\plt  B157 (1985) 179.}
\ldf\FKZL{S.~Ferrara, C.~Kounnas, D.~L\"ust and F.~Zwirner,
CERN preprint CERN-TH-6090-91.}
\ldf\FL{S.\ Ferrara and J. Louis, \plt B278 (1992) 240.}
\ldf\FLT{S.\ Ferrara, D.\ L\"ust and S.\ Theisen,  \plt 242B (1990)
39.}
\ldf\FLST{S.~Ferrara, D.~L\"ust, A.~Shapere and
  S.~Theisen, \plt  B225 (1989) 363;
Y.~Park, M.~Srednicki and A.~Strominger, \plt  B244
(1990) 393.}
\ldf\FMTV{
S. Ferrara, N. Magnoli, T. Taylor and G. Veneziano, \plt
 B245 (1990) 409.}
\ldf\Font{A.\ Font, {\it Periods and Duality Symmetries in \cy\
 Compactifications}, preprint UCVFC/DF-1-92.}
\ldf\FILQ{A.~Font, L.E.~Ib\'a\~nez, D.~L\"ust and F.~Quevedo,
\plt  B245 (1990) 401.}
\ldf\Forsyth{A. Forsyth, {\it Theory of Differential Equations}, Vol.
4,
Dover Publications, New-York (1959).}
\ldf\FSa{S.\ Ferrara and A.\ Strominger, {\it in}  Strings '89,  eds.
R.~Arnowitt, R.~Bryan, M.~Duff, D.~Nanopoulos, C.~Pope, World
Scientific, Singapore, 1989.}
\ldf\FSb{P.\ Fr\`e and P.\ Soriani, {\it Symplectic embeddings,
K\"ahler
geometry and automorphic functions: The Case of $SK(n+1) =
   SU(1,1) / U(1) \times SO(2,n) / SO(2) \times SO(n)$}, preprint
SISSA
 90/91/EP.}
\ldf\GGRS{For a review see
S.~J.~Gates, M.~Grisaru, M.~Ro\v cek and W.~Siegel,
{\it Superspace}, \brk Benjamin/Cummings, 1983.}
\ldf\Griffiths{See, eg., P.\ Griffiths, Ann.\ Math.\ $\us{90}$ (1969)
460.}
\ldf\GS{A.\ Giveon and  D.-J.\ Smit, Mod.~Phys.~Lett. $\us{A6}$
(1991) 2211.}
\ldf\GSW{For a review see M.~Green, J.~Schwarz and E.~Witten,
{\it Superstring Theory}, Cambridge University Press, 1987.}
\ldf\GVW{B. Greene, C.\ Vafa and N.P.\ Warner, \nup324 (1989) 371. }
\ldf\ILR{L.~Ib\~anez, D.~L\"ust and G.~Ross, \plt272 (1991) 251;
       L.~Ib\~anez and D.~L\"ust, CERN preprint CERN-TH.6380/92
(1992).}
\ldf\IN{L. Ib\'a\~nez and H.P. Nilles, \plt  169B
(1986) 354.}
\ldf\kaplunovskya{V. Kaplunovsky, Phys. Rev. Lett  55 (1985)
1036.}
\ldf\kaplunovskyb{V. Kaplunovsky, \nup307 (1988) 145.}
\ldf\kaplunovskyc{V. Kaplunovsky,Texas preprint UTTG-15-91}
\ldf\KL{V. Kaplunovsky and J. Louis,Slac-Pub}
\ldf\KP{C.~Kounnas and M.~Porrati, \plt
   191B (1987) 91.}
\ldf\krasnikov{N.V.~Krasnikov, \plt  193B (1987) 37.}
\ldf\KST{
 A.~Klemm, M.~G.~Schmidt and S.~Theisen,\
{\it Correlation functions
for topological Landau-Ginzburg models with $c\leq3$}, preprint
KA-THEP-91-09.}
\ldf\Kos{B.\ Kostant, Am.\ J.\ Math.\ $\us{81}$ (1959) 973.}
\ldf\lerche{W.\ Lerche,
\nup 238 (1984) 582; W.\ Lerche and W.\ Buchm\"uller,
Ann.\ Phys.\ 175 (1987) 159.}
\ldf\LSW{W.\ Lerche, D.\ Smit and N.\ Warner, \nup372 (1992) 87.}
\ldf\LVW{W.\ Lerche, C.\ Vafa and N.P.\ Warner, \nup324(1989) 427.}
\ldf\louis{J. Louis, SLAC-PUB 5527.}
\ldf\LT{D.~L\"ust and T.~Taylor, \plt  B253 (1991)
335.}
\ldf\maassarani{Z. Maassarani, \plt273B (1991) 457.}
\ldf\MN{G.~Moore and P.~Nelson, Phys. Rev. Lett.   53 (1984) 1519.}
\ldf\morrison{D.~Morrison,
{\it \pf\ equations and mirror maps for hypersurfaces},
 Duke preprint DUK-M-91-14, (1991).}
\ldf\nilles{H.~P.~Nilles, \plt  180B (1986) 240.}
\ldf\NO{H. P.~Nilles and M.~Olechowski, \plt B248 (1990)
268.}
\ldf\NSVZ{V.I.~Novikov, M.A.~Shifman, A.I.~Vainshtein
   and V.I.~Zakharov, \nup260 (1985) 157.}
\ldf\ross{G. Ross, \plt  211B (1988) 315.}
\ldf\seiberg{N.~Seiberg, \nup303 (1988) 206.}
\ldf\shenker{S.H.~Shenker, Lecture at
1990 Carg\`ese Workshop, Rutgers preprint RU-90-47;
A.~Dabholkar, \nup368 (1992) 293.}
\ldf\SV{M.A.~Shifman and A.I.~Vainshtein,
   \nup277 (1986) 456, \nup359 (1991) 571.}
\ldf\strominger{A.\ Strominger, \cmp133 (1990) 163.}
\ldf\taylora{T.R.~Taylor, \plt  164B (1985) 43.}
\ldf\taylorb{T.~Taylor, \plt  B252 (1990) 59.}
\ldf\TV{T. R.~Taylor and G.~Veneziano, \plt  212B
(1988) 147.}
\ldf\vafa{C.\ Vafa, Mod.\ Phys.\ Lett. A6 (1991) 337.}
\ldf\VW{E.\ Verlinde and N.P.\ Warner, \plt 269B (1991) 96.}
\ldf\weinberg{S.~Weinberg, Phys. Lett B91 (1980) 51.}
\ldf\witten{E.~Witten, \plt  149B (1984) 351
   , \plt  155B (1985) 151.}
\ldf\zwiebach{See for example B.~Zwiebach, MIT preprint MIT-CTP-1926
and references therein.}
%
%
%
%
%
\ldf\cubicF{E.\ Cremmer, C.\ Kounnas, A.\ Van Proeyen, J.\
Derendinger,
S.\ Ferrara, B.\ de Wit and L.\ Girardello, \nup250 (1985) 385 \semi
   other cubic F-theories, de Wit xxxxxx?????}
\ldf\dfnt{B.\ de Wit and A.\ Van Proeyen, \nup245 (1984) 89;
B.\ de Wit, P.\ Lauwers and A.\ Van Proeyen, \nup255 (1985) 569;
E.\ Cremmer, C.\ Kounnas, A. \ Van Proeyen, J.P.\ Derendinger,
S.\ Ferrara, B.\ de Wit and L.\ Girardello,  \nup250 (1985) 385.
}
\ldf\fetal{L.~Castellani, R.~D'Auria and S.~Ferrara,
    \plt B241 (1990) 57; Class.\ Quant.\ Grav.\ 7 (1990) 1767;
R.~D'Auria,
S.~Ferrara and P.~Fr\'e, \nup359 (1991) 705.}
\ldf\cubicF{E.~Cremmer and A.~Van Proeyen, Class. Quant. Grav.
{\bf 2} (1985) 445;
S.\ Cecotti, \cmp124(1989) 23;
B.~de Wit and A.~Van Proeyen, {\it Special geometry, cubic
polynomials
and homogeneous quaternionic spaces},
CERN--preprint  TH.6302/91.}
\ldf\typeII{??xx}
 %
%
%
\ldf\tntwo{E.\ Witten, \cmp118 (1988) 411, \nup340 (1990) 281; T.\
Eguchi and
 S.K.\ Yang, Mod.\ Phys.\ Lett.\ $\us{A5}$ (1990) 1693.}
\ldf\AMW{P.\ Aspinwall and D.\ Morrison, Duke
preprint DUK-M-91-12; Contributions of C.~Vafa and
E.\ Witten {\sl in} {\it Essays on Mirror Manifolds}, ed.
S.-T.~Yau, 1992, International Press, Hong Kong.}
\ldf\PFref{
A.\ Cadavid and S.\ Ferrara, \plt B267 (1991) 193;
W.\ Lerche, D.\ Smit and N.\ Warner, \nup372 (1992) 87;
D.~Morrison,
 Duke preprint DUK-M-91-14;
S.~Ferrara and J.~Louis, \plt 278B (1992) 240;
A.~Ceresole, R.~D'Auria, S.~Ferrara, W.~Lerche and J.~Louis,
CERN preprint CERN-TH.6441/92.}
\ldf\top{B. Blok and  A.\ Varchenko, Int.J.Mod.Phys.$\us{A7}$, (1992)
1467; E.\ Verlinde and N.P.\ Warner, \plt B269 (1991) 96; Z.
Maassarani, \plt B273 (1991) 457; A.~Klemm, M.~G.~Schmidt and
S.~Theisen,\
{\it Correlation functions
for topological Landau-Ginzburg models with $c\leq3$}, preprint
KA-THEP-91-09; P.\ Fr\'e, L.\ Girardello, A.\ Lerda, and P.\
Soriani, {\it Topological First-Order Systems with \LG\ Interactions},
 Trieste preprint SISSA/92/EP (1992).}
%
%
%
\ldf\Ferrara{S.\ Cecotti, \cmp131(1990) 517;
A.\ Cadavid, M.\ Bodner and S.\ Ferrara, \plt B247 (1991) 25.}
\ldf\FKT{A.\ Font, {\it Periods and Duality Symmetries in \cy\
Compactifications}, preprint UCVFC/DF-1-92; A.~Klemm and S.~Theisen,
{\it Considerations of one modulus Calabi-Yau
   compactifications: Picard-Fuchs equations, K\"ahler potentials and
mirror maps}, Karlsruhe preprint
   KA-THEP-03-92.}
%
%
%
\ldf\mirror{L.\ Dixon and D.\ Gepner, unpublished;
W.\ Lerche, C.\ Vafa and N.P.\ Warner, \nup324(1989) 427;
B.\ Greene and M.\ Plesser, \nup 338 (1990) 15;
P.\ Candelas, M.\ Lynker and R.\ Schimmrigk, \nup 341 (1990) 383.}
\ldf\mirrorreview{For a review see
B.\ Greene and M.\ Plesser, {\sl in} {\it Essays on Mirror
Manifolds}, ed.
S.-T.~Yau, 1992, International Press, Hong Kong.}
\ldf\twopara{Candelas, de la Ossa, Font, Schimmrigk??xx}
%
%
%
\ldf\gauginoc{For a review see for example H.~P.~Nilles,
Int. J. Mod. Phys. $\us{A5}$ (1990) 4199 and references
therein.}
\ldf\amati{D. Amati, K. Konishi, Y. Meurice, G. Rossi and
G. Veneziano, Phys. Rep.  162 (1988) 169.}
\ldf\dindrsw{J.P.~Derendinger, L.E.~Ib\'a\~nez and
  H.P.~Nilles, \plt  155B (1985) 65;
M.~Dine, R.~Rohm, N.~Seiberg and
  E.~Witten, \plt  156B (1985) 55.}
\ldf\effl{
A.~Font, L.E.~Ib\'a\~nez, D.~L\"ust and F.~Quevedo,
\plt  B245 (1990) 401;
S.~Ferrara, N.~Magnoli, T.~Taylor and G.~Veneziano, \plt
 B245 (1990) 409;
H.~P.~Nilles and M.~Olechowski, \plt B248 (1990)
268;
P.~Bin\'etruy and M. K. Gaillard, \nup358 (1991) 121;
M.~Cveti\v c, A.~Font, L.E.~Ib\'a\~nez, D.~L\"ust and
F.~Quevedo, \nup361 (1991) 194.}
\ldf\TVY{G.~Veneziano and S.~Yankielowicz, \plt
   113B (1982) 231;
  T.R.~Taylor, G. Veneziano and S.~Yankielowicz, \nup
   B218 (1983) 493;
T.R.~Taylor, \plt  164B (1985) 43.}
\ldf\hiddenmatter{D.~L\"ust and T.~Taylor, \plt B253 (1991)
335;
B.~de Carlos, J.~A.~Casas and C.~Mu\~noz, Phys. Lett B263 (1991)
248, CERN preprint CERN-TH.6436/92.}
\ldf\twogaugino{L.~Dixon, talk presented at the A.P.S. D.P.F.
  Meeting, Houston, 1990, SLAC-PUB 5229;
J.~A.~Casas, Z.~Lalak, C.~Mu\~noz and G.G.~Ross, \nup347 (1990) 243;
T.~Taylor, \plt B252 (1990) 59;
L.~Dixon, V.~Kaplunovsky, J.~Louis and M.~Peskin, unpublished.}
%
%
%
\ldf\wnrt{E.~Martinec, \plt 171B (1986) 189;
 M.~Dine and N.~Seiberg, \prl 57 (1986) 2625;
 J.~Atick, G.~Moore and A.~Sen, \nup307 (1988) 221;
 O.~Lechtenfeld and W.~Lerche, \plt 227B (1989) 375.}
\ldf\JJW{I.~Jack and D.~R.~T.~Jones, \plt258 (1991) 382;
P.~West, \plt258 (1991) 375.}
\ldf\frenormalization{M.A.~Shifman and A.I.~Vainshtein,
   \nup277 (1986) 456;
H.~P.~Nilles, \plt  180B (1986) 240;
I.~Antoniadis, K.~Narain and T.~Taylor, \plt B267 (1991) 37.}
%
%
%
%
\ldf\threegen{See for example
B.~Greene, A.~Lutken and G.~Ross, \nup325 (1989) 101;
G.~Ross, CERN preprint TH-5109/88 and references therein.}
%
%
%
\ldf\ellisetal{For a review see J.~L.~Lopez and D.~V.~Nanopoulos,
Texas preprint CTP-TAMU-76/91 and references therein.}
%
%
%
\ldf\danomaly{
J.~Louis, SLAC-PUB 5527;
J.~P.~Derendinger, S.~Ferrara, C.~Kounnas and F.~Zwirner,
CERN preprint CERN-TH.6004/91-REV;
G.~Cardoso Lopes and B.~Ovrut, \nup369 (1992) 351;
V.~Kaplunovsky and J.~Louis, SLAC-PUB to appear.}
\ldf\ILRE{I.~Antoniadis, J.~Ellis, S.~Kelley and D.~V.~Nanopoulos,
\plt B272 (1991) 31;
L.~Ib\'a\~nez, D.~L\"ust and G.~Ross, \plt B272 (1991) 251;
       L.~Ib\'a\~nez
       and D.~L\"ust, CERN preprint CERN-TH.6380/92 (1992).}
\ldf\DKLANT{L.~Dixon, V.~Kaplunovsky and J.~Louis, \nup355 (1991)
649;
I.~Antoniadis, K.~Narain and T.~Taylor, \plt B267 (1991) 37.}
%
%
%
\ldf\schell{See {\it Superstring Construction}, Current Physics
  Sources and Comments, Vol. IV, ed. A.~Schellekens
  (North-Holland, 1989).}
\ldf\twotworeview{For a review see
L. Dixon, {\sl in} Proc. of the 1987 ICTP Summer
Workshop in High Energy Physics, Trieste, Italy, eds.~G.~Furlan et
al.;
D.~Gepner, {\sl in} Proc. of the 1989 ICTP Spring School on
Superstrings,
Trieste, Italy, eds.~M.~Green et al.;
B.~Greene, Lectures at the ITCP
Summer School in High Energy Physics and Cosmology, Trieste, Italy,
1990; J.~Distler, these proceedings,
and references therein.}
\ldf\cargese{See for example Proceedings of Carg\`ese Workshop on
Random Surfaces, Quantum Gravity and Strings, Carg\`ese, May 1990.}
\ldf\ntworeview{C.~Vafa, {\it in}    Proceedings of Trieste Summer
School on High Energy Physics and
   Cosmology, Trieste 1989; S.\ Cecotti,
 \nup 355 (1991) 755; Int.\ J.\ Mod.\ Phys.\ $\us{A6}$ (1991) 1749.}
%
%
%
\ldf\dualityreview{For a review see S.~Ferrara and S.~Theisen, {\sl
in}
Proc. of the Hellenic Summer School 1989, World Scientific;
D.~L\"ust, CERN preprint
CERN-TH.6143/91;
J.~Erler, D.~Jungnickel, H.~P.~Nilles amd M.~Spali\'nski, MPI
preprint
MPI-Ph/91-104
and references therein.}
%
%
%
\ldf\BH{See for example
Proceedings of the  Spring School on String Theory and Quantum
Gravity, ITCP, Trieste, April 1992.}
%
%
%
\ldf\CFGP{E.~Cremmer, S.~Ferrara, L.~Girardello and A.~Van Proeyen,
\nup212 (1983) 413.}
%
%
%
\ldf\tpcecotti{S.~Cecotti, these proceedings.}
\ldf\tpcandelas{P.~Candelas, these proceedings.}
%
%
\ldf\kln{S.~Kalara, J.~Lopez and D.~Nanopoulos, preprint
CTP-TAMU-46-91.}
\ldf\howi{Y. Hosotani, \plt  126B (1983) 309;\brk
 E. Witten, \plt  126B (1984) 351.}%
\ldf\thooft{G.~`t~Hooft, \nup72 (74) 461.}
\ldf\aeln{I.~Antoniadis, J.~Ellis, A.B.~Lahanas and
  D.V.~Nanopoulos, preprint CERN-TH.5604/89.}
\ldf\gs{
  J.~Scherk and J.H.~Schwarz, \nup81 (1974) 118;\brk
  D.~Gross and J.~Sloan, \nup291 (1987) 41.}
\ldf\bfq{C.~Burgess, A.~Font and F.~Quevedo, Nucl.
   Phys.  B272 (1986) 661.}
\ldf\ginsparg{P.~Ginsparg, \plt  197B (1987) 139.}
\ldf\ellis{
J.~Ellis, P.~Jetzer and L.~Mizrachi, \plt  196B (1987) 492
, \nup303 (1988) 1.}
\ldf\wb{J.~Wess and J.~Bagger, {\it Supersymmetry and
  Supergravity} (Princeton Unversity Press, 1983);\brk
and preprint JHU-TIPAC-9009, June 90.}
\ldf\PrandT{J.~Preskill and S.~Trivedi,
  \nup (Proc. Suppl.)  1A (1987) 83.}
\ldf\witind{E.~Witten, \nup202 (1982) 253.}
 \ldf\rv{G.C.~Rossi and G.~Veneziano, \plt  138B
    (1984) 195.}
\ldf\grisaru{M.T.~Grisaru, B.~Milewski and D.~Zanon,
  in {\it Supersymmetry and Its Applications -- Superstrings,
  Anomalies and Supergravity}, eds. G.W.~Gibbons, S.W.~Hawking and
  P.K.~Townsend (Cambridge Univ. Press, 1986).}
\ldf\cdg{R.~Jackiw and C.~Rebbi, Phys. Rev. Lett.  37 (76)
   132;\brk
  C.G.~Callan, R.F.~Dashen and D.J.~Gross,
   \plt  63B (76) 334.}
\ldf\serre{see for example J.-P. Serre, {\sl A Course in Arithmetic},
Springer-Verlag 1973, New York.}
\ldf\jones{D.R.T.~Jones, \nup87 (75) 127.}
\ldf\konishi{T.E.~Clark, O.~Piquet and K.~Sibold, \nup
   B159 (79) 1;\brk
   K.~Konishi, \plt  135B (1984) 439.}
\ldf\bganom{P.~Bin\'etruy and M.K.~Gaillard, \plt
   232B (1989) 83.}
\ldf\bq{C.P.~Burgess and F.~Quevedo, Phy. Rev. Lett. 64 (1990)
2611.}
%
\def\brk{\hfill\break}
\def\LG{Lan\-dau--Ginz\-burg}
\def\cy{Calabi--Yau}
\def\K{K\"ahler}
\def\pf{Picard--Fuchs}

\def\sc{SCFT}

\def\leel{low-energy effective Lagrangian}
\def\el{effective Lagrangian}

\def\del{\partial}

\def\del{\partial}

\def\Db{\overline{D}}
\def\zb{\overline{z}}

\def\fb{\overline{f}}

\def\cy{Calabi--Yau}

\def\Kh{\hat{K}}
\def\Knh{{\cal K}}

\def\V{{\cal V}}
\def\T{T}

\def\Gammah{\hat{\Gamma}}

\def\cS{{\cal K}}
\def\IE{\relax{{\rm I\kern-.18em E}}}

\def\IGam{\relax{{\rm I}\kern-.18em \Gamma}}

\def\LG{Lan\-dau-Ginz\-burg\ }

\def\WABC{W_{\a\b\c}}
\def\W{{\cal W}}

\def\bv{{\bf V}}

\def\sc{SCFT}
\def\leel{low energy effective Lagrangian}
\def\pf{Picard--Fuchs}
\def\el{effective Lagrangian}

\def\Db{\overline{D}}
\def\zb{\overline{z}}

\def\fb{\overline{f}}

\def\cy{Calabi--Yau}

\def\Kh{\hat{K}}
\def\Knh{{\cal K}}

\def\V{{V}}
\def\T{T}
\def\Gammah{\hat{\Gamma}}

\def\K{K\"ahler}

\def\w{w}
%
%
\def\cernout{
\def\pubnum{\hbox{CERN-TH.6580/92}}
\def\pdate{
\hbox{CERN-TH.6580/92}
\hbox{July 1992}
}
\titlepage
\vskip 2cm
\title{Differential Equations in Special \K\ Geometry}
\author{Jan Louis}
\CERN
\vskip 1.5cm
\abstract{
The structure of differential equations as they appear in   special
\K\ geometry of $N=2$ supergravity and $(2,2)$ vacua of the heterotic
string is summarized.  Their use for computing couplings in
the low energy effective Lagrangians of string compactifications is
outlined. }
\vskip 3cm
\centerline
{\it Talk presented at the Workshop on String Theory, April 8--10,
1992, Trieste, Italy}
         }
%

\cernout
\endpage

\chap{Introduction}

In order to test string theory as a theory of unifying all known
interactions one needs to extract its low-energy limit and compare it
with the Standard Model. A first step in this program is the
derivation of a
low energy effective Lagrangian which only depends on the massless
string modes.
The process of ``integrating out'' the heavy string modes
depends on the string vacuum
chosen  and thus leads to a different \el\ for each vacuum.
Phenomenological prejudice focuses our attention on vacua displaying
 $N=1$ supersymmetry in four space--time dimensions. This subclass of
string vacua can be characterized by
 an $N=2,c=9$ worldsheet super conformal field theory (SCFT) in the
  left--moving
sector \BDFS. Furthermore, the couplings of the corresponding \leel\
are
directly
related to correlation functions in the \sc. Unfortunately, for
most
string vacua we are currently not able to calculate the relevant
correlation
functions. However, recently for a particular family of
string
vacua (a compactification on a specific \cy\ threefold), some of the
couplings
in the \leel\ have been computed exactly (at the string tree-level,
but to all
orders in the $\sigma$--model coupling) by using techniques of
algebraic geometry and without ever relying on the underlying \sc\
\CDGP. It was shown that the couplings
could be obtained from the solution of a certain 4th-order linear
holomorphic differential equation. Subsequently, it was realized that
this
differential equation is a particular case of the so-called ``\pf\
equations'' obeyed by the
periods of the holomorphic three-form $\Omega$ that exists on any
\cy\
threefold \nrf{\CF\LSW\morrison} \refs{\CF{--}\morrison}.\foot{
\pf\ equations can be derived for general
``\cy'' $d$--folds \doubref\LSW\CF, but we consider only $d=3$ in the
following.}
A further step in uncovering the general structure behind the
differential
equation was undertaken in refs.~\refs{\FL,\CDFLL}. It was shown that
the \pf\ equations
for a \cy\ threefold are just another way of expressing a geometrical
structure
called ``special \K\ geometry''
\nrf{\dfnt\CFG\FSa\DKLa\Ferrara\strominger\fetal\CD}
\refs{\dfnt{--}\CD}.

Special \K\ geometry
first arose in the study of coupling vector
multiplets
to $N=2$ supergravity in four dimensions \dfnt. The manifold spanned
by the scalars of the vector multiplets turned out to be a \K\
manifold with an additional constraint dictated by  $N=2$
supergravity.\foot{
A coordinate--free characterization of special geometry was
given in
\fetal\ in the context of $N=2$ supergravity and in
\refs{\strominger{,}\CD}\ for a
Calabi-Yau moduli space.}
Consequently, the same structure
also appears in appropriate compactifications of type II string
theories \refs{\seiberg,\CFG}.
However, it can also arise in $N=1$ vacua of the heterotic string if
they display an additional right-moving $N=2$ worldsheet
supersymmetry (so-called $(2,2)$ vacua)
\refs{\seiberg,\CFG,\DKLa}.
In this case
the couplings of the  $N=1$ effective Lagrangian  obey the
constraints of special geometry; the
\K\ potential $K$ (which encodes the kinetic terms of the scalar
fields) is related to
 the (holomorphic) Yukawa couplings $\WABC$. Both quantities can
be expressed in terms of so-called holomorphic prepotentials
$X^A, F_A$.
Therefore, in $(2,2)$ vacua
it is sufficient to compute  the prepotentials in order to determine
the couplings of the \el\ and
 the differential equation of
ref.~\CDGP\ is precisely an equation which determines $X^A$ and
$F_A$.

The content of ref.~\CDGP\ is reviewed by P.~Candelas in these
proceedings. In this talk we discuss  the \pf\
equations
from the point of view of special \K\ geometry following
refs.~\refs{\FL,\CDFLL}. This analysis clarifies the exact relation
of the \pf\ equations to the couplings of an \el\ for $(2,2)$
heterotic vacua.
 The organisation   is as follows. In
section~2,
we briefly recall the basic definitions and properties of special
geometry as they arise in $N=2$ supergravity and $(2,2)$ vacua of
the heterotic string. In section~3 we give an entirely
equivalent formulation of this geometrical structure which features
a purely holomorphic differential identity. The rest of the talk then
evolves around this holomorphic identity. In order to make contact
with ref.~\CDGP\ the special case
of a one-dimensional \K\ manifold is treated in section~4. Here, we
also observe further properties of the holomorphic identity derived
in section~3.  The moduli space of
\cy\ threefolds  is a subclass of special \K\ manifolds
\refs{\FSa,\strominger,\CD} and so we briefly relate the discussion
of section~3 to  this subclass in section~5.
The holomorphic differential identity of section~3 corresponds to
the \pf\ equations obeyed by the periods on the \cy\ manifold.
The important point is that
the coefficient functions of the differential identity
can be computed from the defining polynomial of
the \cy\ manifold or equivalently from the \LG\ superpotential which
characterizes the string vacuum. Once the coefficients are known, the
identity  turns into a non-trivial linear differential equation whose
solutions are the prepotentials $X^A$ and $F_A$ which also determine
$\WABC$ and $K$. This indicates that
it  not always necessary
to calculate the correlation functions of the SCFT. Instead,
the low energy
couplings are already encoded in the \LG\ superpotential. In
section~6 we outline how the coefficients of the differential
identity can be  computed from the \LG\ superpotential.

Special geometry also made its appearance in the context of
topological conformal field theories (TCFT) \refs{\DVV,\CV} and
differential equations similar to the \pf\ equations
govern the space of topological deformations. In fact, these
observations were one motivation for the investigation performed in
refs.~\refs{\FL,\CDFLL}.
In section 7 we conclude by  relating special geometry as it arises
in TCFT to the analysis of the previous sections.

This talk is based on refs.~\FL\ and \CDFLL\ where one also finds a
lot of the technical details omitted here.

\chap{ Special \K\ Geometry}

Let us first briefly summarize the definition and some of the basic
properties of special \K\ geometry \refs{\dfnt,\fetal}.

The metric of
an $n$--dimensional \K\ manifold $\cM$ is given by
$$
g_{\a{\bar \b}}(z,\zb)=\del_\a \del_{\bar \b}K(z,\zb)  \, , \qquad \a
= 1,\ldots,n \, ,
\eqn\metric
$$
where the \K\ potential $K(z,\zb)$ is a real function of the complex
coordinates $z^\alpha$ and
$\zb^{\bar \alpha}$.

Special \K\ geometry is defined by an additional constraint on the
K\"ahler
        potential which reads
$$
K(z,\zb) = -\ln i\left( X^A(z)\bar F_A(\zb)-\bar X^A (\zb) F_A
(z)\right) \, , \qquad  A=0,\ldots, n  \, ,
  \eqn\kdef
$$
where $X^A(z)$ is  holomorphic  and $F(X^A)$ is
a holomorphic functional  homogeneous of degree
$2$
in  $X^A$:
$$
\del_{\bar \a}
X^A=0 \, , \qquad
2F=X^A F_A(X)\, , \qquad F_A\equiv \coeff{\del F}{\del X^A}\, .
\eqn\hom
$$
Let us note that $K$ is expressed in
terms of the purely holomorphic objects $X^A(z)$ and $F(X^A)$ and
their complex conjugates.
 The metric $g_{\alpha \bar \beta}$ as defined in eq.~\metric\
is invariant under the K\"ahler
transformations
$K(z,\zb) \rightarrow  K(z,\zb) + f(z) + \fb (\zb)$. For the \K\
potential \kdef\ this translates into the transformation properties
$$
X^A \rightarrow X^A e^{-f}\, , \qquad F_A \rightarrow F_A e^{-f} \, .
\eqn\ktransxf
$$

It also proves useful to introduce the $(2n+2)$ dimensional row
vector\foot
{We take the expression $(X^A, F_A)$ always as
an abbreviation for $(X^0, X^a, F_a, -F_0)$.}
$$
V\  =\ (X^A, F_A)\ :\equiv\ (X^0,X^\a,F_\a,-F_0)\ .
\eqn\sive
$$
In terms of $V$ we find
$$
K = - \ln\left(V(-iQ) V^\dagger \right) \, ,
\eqn\ksymp
$$
where $Q$ is a
symplectic metric which satisfies $Q^2=-1\, ,Q=-Q^T$ and reads
$$
 Q\ =\ \pmatrix{
  & & & 1 \cr
  & & -\bfone_n & \cr
  & \bfone_n & & \cr
  -1 & & &\cr}
 \ . \eqn\Qsympdef
$$
However, $V$ is not uniquely defined. Under a global $Sp(2n+2)$
rotation one finds \refs{\dfnt{,}\CFG{,}\fetal,\CD}
$$
\big(\tilde X^A , \tilde F_A(\tilde X^A)\big)\ =\
\big(X^A, F_A(X^A)\big)\cdot M\ ,\qquad M\in Sp(2n+2)\ ,
\eqn\spaction
$$
where $\tilde F_A = (\del \tilde F/\del \tilde X^A) $ and $\tilde F$
is
again a homogeneous function of $\tilde{X}^A$ of degree 2. From
eq.~\ksymp\ we see
that $K$ is manifestly invariant under such reparametrizations.

As a consequence of eqs.~\kdef\ and \hom\ $V$ satisfies the following
set of covariant identities
$$\eqalign{
D_\a V\ &=\ U_\a \, ,\cr
D_\a U_\b\ &=\ - i C_{\a\b\c}g^{\c \bar\d} \bar{U}_{\bar\d} \, ,\cr
D_\a {\bar U}_{\bar \b}\ &=\ g_{\a\bar \b} {\bar V}\, , \cr
D_\a \bar V\ &=\ 0  \,  . }
\eqn\sis
$$
The first equation is the definition of $U_\a$ and the \K\ covariant
derivative $D_\a$ has been defined as
follows
$$\eqalign{
D_\a  V = (\del_\a + \del_\a K) V \, ,\qquad  D_\a \bar V = \del_\a
\bar V \, , \cr
D_\a U_\b\ =\ (\del_\a +\del_\a K)U_\b - \Gamma_{\a\b}^\c U_\c \
\, ,\qquad
D_{ \a}{\bar U}_{\bar \b}\ =\ \del_{ \a}{\bar U}_{\bar \b}\, .
}
\eqn\der
$$
Here $\Gamma_{\a\b}^\c$ ($= g^{\a\bar{\d}} \del_\a g_{\bar{\d}\b}$)
denotes the usual Christoffel connection of the K\"ahler manifold and
$\del_\alpha K, \del_{\bar \alpha}K$ act as connections for K\"ahler
transformations \ktransxf.
($\del_\alpha K$ is an Abelian
 connection
of a holomorphic line bundle $L$ whose first Chern class is the
K\"ahler class.)
Finally, we abbreviate
$$
C_{\a\b\c} =
e^K W_{\a\b\c}\, , \qquad W_{\a\b\c} =
\del_\a
X^A \del_\b X^B \del_\c X^C F_{ABC}   \, ,
\eqn\fabc
$$
where $W_{\a\b\c}$ is holomorphic: $\del_{\bar \a} W_{\a\b\c}$.
As for eq.~\sis\ one derives a set of equations including the
anti--holomorphic
derivative $D_{\bar \alpha}$.

As an  integrability condition of the second equation in \sis\ one
finds
$$
R_{\bar\a \b  \d}^\c\ =\
 g_{\bar \a \b}\d_\d^\c
+g_{\bar\a\d}\d_\b^\c-
 C_{\b \d \mu}g^{\mu \bar\mu}
C_{\bar \mu \bar \a \bar \c} g^{\c \bar\c}  \, ,
\eqn\curvb
$$
where $R_{\bar\a \b  \d}^\c$ ($= \del_{\bar \a} \Gamma_{\b\d}^\c$)
denotes the Riemann tensor of the Christoffel connection.
The Bianchi identities then imply
$$
\eqalign{
\Db_{\bar \epsilon} C_{\alpha\beta\gamma} =&\ 0\, ,\cr
D_{\epsilon} C_{\alpha\beta\gamma} -
D_{\alpha} C_{\epsilon\beta\gamma} =&\ 0 \, .}
\eqn\dc
$$
In $(2,2)$ vacua of the heterotic string
eqs.~\curvb\ and \dc\ were shown to arise from Ward identities of the
right-moving $N=2$ world-sheet supersymmetry  \DKLa. In such
theories the scalar fields $z^\a$ correspond to the so-called moduli
fields of the string spectrum and
$C_{\alpha\beta\gamma}$ are the (moduli dependent) Yukawa couplings
of the ${\bf 27}, ({\bf \bar{27}})$ matter multiplets.

\chap{Holomorphic Differential Identities}
So far we recapitulated the basic definitions of special geometry. In
this section we show that there is  another way of expressing
the same constraints.
Starting from eqs.~\sis\ we will see that it is
possible to
derive an equivalent but completely holomorphic set of identities
which manifestly display the fact that special
geometry is determined entirely in terms of the holomorphic sections
$X^A, F_A$.
It is these holomorphic identities which can be used to compute
the couplings of the \el\ from the \LG\ superpotential. Furthermore,
they allow us to make contact with the \pf\ equations of
the \cy\ manifold and the corresponding equations of TCFT.

Let us first observe that
eq.~\sis\ can be written very compactly as a $(2n+2)\times(2n+2)$
matrix equation
$$
(\bfone{\del_\a} - {\cA_\a}) {\bf U}\ =\ 0\ ,
\eqn\linearsi
$$
where ${\bf U}=(V,U_\a,\bar{U}_{\bar \a},\bar{V})^T$ and
$$
\cA_\a\ =\ \pmatrix{
-\del_\a K & \delta_\a^\b & 0 & 0 \cr
0 &- \delta_\c^\b \del_\a K + \Gamma _{\c\a}^\b &
- i C_{\a\b\c}g^{\c \bar{\c}}& 0\cr
0 & 0 & 0 & g_{\a\bar{\b}} \cr
 0 & 0 & 0 & 0 } \ .
\eqn\gammanonhol
$$
{}From eq.~\sis\ one also infers that in addition ${\bf U}$
satisfies
$$
 (\bfone{\del_{\bar\a}} - {\cA_{\bar\a}})
{\bf U}\ =\ 0\ ,
\eqn\complexco
$$
where
$$
\cA_{\bar\a}\ = \ \pmatrix{
0 & 0 & 0 & 0 \cr
g_{\bar{\a}\b} & 0 & 0 & 0 \cr
0 & i C_{\bar\a\bar\b\bar\c} g^{\bar\c \c }
& -\delta _{\bar \c} ^ {\bar \b} \del_{\bar
\a} K +\Gamma_{\bar\c\bar\a}^{\bar \b} & 0 \cr
0 & 0 & \delta_{\bar{\a}}^{\bar{\b}} & -\del_{\bar\a} K }\ .
\eqn\gacomplexco
$$
Strominger observed that
as a consequence of \curvb\ and \fabc\ the connections
$\cA_\a$ and ${\cA}_{\bar\a}$ have vanishing curvature \strominger:
$$
F_{\a\b} = F_{\bar\a\bar\b} = F_{\a\bar\b}\ =\ 0\, , \qquad F_{\a\b}
= \del_{[\a} \cA_{\b]} -  [\cA_\a, \cA_\b ]\, .
\eqn\zerocu
$$
This zero curvature condition is yet another characterization of
special
\K\ geometry.

{}From the definition of the covariant derivatives in eq.~\der\
we see that the set of identities \sis\ (or equivalently \linearsi)
is covariant with respect to \K\ and coordinate transformations.
However, eqs.~\linearsi\ and \complexco\ are also
covariant under the more general
 gauge transformations
$$
{\bf U}' = S^{-1}\cdot {\bf U}\, ,\qquad \cA_\a' = S^{-1}\cA_\a
S-S^{-1}\del_\a S\, ,\qquad \cA_{\bar \a}' = S^{-1}\cA_{\bar \a}
S-S^{-1}\del_{\bar \a} S\, .
\eqn\gauget
$$
The important point is that via a non-holomorphic transformation
of the form
$$
S\ =\ \pmatrix{ \ast_{1\times 1} & {\bf 0} & {\bf 0} & 0\cr
               \ast & \ast_{n\times n} & {\bf 0} & {\bf 0} \cr
               \ast & \ast & \ast_{n\times n} & {\bf 0} \cr
                \ast & \ast & \ast & \ast_{1 \times 1}}\ \in B,
\eqn\gaugetrans
$$
($B$ denotes the Borel subgroup   of $SL(2n+2,\IC)$) one can gauge
away $\cA_{\bar\a}$ completely and simultaneously
 make $\cA_\a$ purely holomorphic:
 $$
{\cA}_{\bar\a}' = 0\, , \qquad
\del_{\bar\a} A_\a=0 \, ,\qquad
{\bf V}=  S\, {\bf U}\ ,\qquad
\del_{\bar\a}{\bf V}=0 \, ,
\eqn\puregauge
$$
(where we have denoted the holomorphic quantities by $A_\a$ and $\bv$
respectively).  In this holomorphic gauge
 eqs.~\linearsi, \complexco\ are replaced by
$$
(\bfone{\del_\a} - A_\a) {\bf V}\
=\ 0\ ,
\eqn\dholo
$$
which is an entirely equivalent characterization of special geometry.
It is this holomorphic form which
will be the focus of interest in the rest of this talk. Here we note
that the condition \puregauge\ does not completely fix the gauge
freedom \gauget. Eq.~\dholo\
still displays a residual gauge symmetry of purely holomorphic
$S$-transformations.

Let us assemble a few more properties of
$A_\a$. It is not an arbitrary
$(2n+2)\times (2n+2)$ matrix but can be brought to  the form (by
using  holomorphic $S$-transformations)
$$
A_\a\ =\  \IGam_\a + \IC_\a\ ,
\eqn\ac
$$
where
$$
\IGam_\a
 =\
 \pmatrix{
-\del_\a\Kh & 0& 0 & 0 \cr
0 & (\Gammah_\a-\del_\a\Kh\bfone)_{\b}^\c & 0 & 0\cr
0 & 0 & (\del_\a\Kh\bfone-\Gammah_\a)^\b_{\c} & 0 \cr
0 & 0 & 0 & \del_\a\Kh\cr}\ ,
\eqn\gammahat
$$
and
$$
\IC_\a
 =\ \pmatrix{
0& \d_\a^\c & 0 & 0 \cr
0 & 0& (W_\a)_{\c\b} & 0\cr
0 & 0 & 0 & \d^\b_\a \cr
0 & 0 & 0 & 0}\ .
\eqn\cdef
$$
In order to define the
 ``hatted'' connection in eq.~\gammahat\ let us first introduce
$$
t^a(z)={{X^a}(z)\over{X^0}(z)} \, , \qquad a= 1, \ldots,n \, .
\eqn\speco
$$
In terms of $t^a$ and $X^0$ one finds  that the \K\ connection
$\del_\a K$, as well as the Christoffel connection
$\Gamma_{\alpha\beta}^\gamma$, split into a holomorphic piece
($\Kh_\alpha (z)$ and $\Gammah_{\alpha\beta}^\gamma (z)$) which still
transforms as a connection, and a non-holomorphic piece with tensorial
transformation properties:
$$
\eqalign{
\del_\alpha K(z,\zb) =\ &
\Kh_\alpha (z) +
\Knh_\alpha (z,\zb)  \, ,  \cr
\Gamma_{\alpha\beta}^\gamma (z,\zb) =\ &
\Gammah_{\alpha\beta}^\gamma (z) +
\T_{\alpha\beta}^\gamma (z,\zb) \, . }
\eqn\conndef
$$
where
$$
\eqalign{
\Kh_\alpha (z) =& - \del_\alpha \ln X^0 (z)\, ,\cr
\Gammah_{\alpha\beta}^\gamma (z) =&\ (\del_\beta e_\alpha^a)
 e^{-1 \gamma}_a \, , \cr
 e_\alpha^a(z) =&\ \del_\alpha t^a (z)    \, .
}
\eqn\tconndef
$$
(The expressions for $\Knh_\alpha$ and $\T_{\alpha\beta}^\gamma$ can
be found in ref.~\FL, they are not essential in the following.) The
important point is that
the holomorphic objects  ${\hat K_\a}$ and
 $\hat \Gamma^\a_{\b\c}$ transform as connections
under  \K\ and holomorphic reparametrizations respectively;
moreover $T^\a_{\b\c}$ is a tensor under holomorphic diffeomorphisms
and  $\cK_\a$ is \K\ invariant.
As a consequence one can define
 holomorphic covariant derivatives
in analogy with eq.~\der\ where all connections are replaced by
their hatted analogue. From eq.~\gammahat\ we see that these
holomorphic covariant derivatives exactly appear in \dholo.
It is easy to check that $\hat \Gamma^\a_{\b\c}$ is also flat:
$$
{\hat R}^\c_{\d\a\b}
\equiv \del_\d{\hat\Gamma}^\c_{\a\b}-\del_\a{\hat \Gamma}
^\c_{\d\b}+{\hat \Gamma}^\mu_{\a\b}{\hat
\Gamma}^\c_{\mu\d}-{\hat\Gamma}^\mu_{\d\b}{\hat
\Gamma}^\c_{\mu\a}=0\, .
\eqn\hatcurva
$$
The flat coordinates are the so-called ``special coordinates''
$t^a=z^\a$.
In these coordinates we find from \tconndef
$$
e^a_\a=\delta_\a^a, \qquad {\hat \Gamma}^\d_{\a\b}=0.
\eqn\flutstuff
$$
The \K\ gauge choice $X^0=1$ implies $\hat K_\a=0$. From
eq.~\gammahat\
we learn that in these coordinates also $\IGam_\a = 0$ holds whereas
eq.~\fabc\ shows
$$
 W_{\a\b\c} = \del_\a \del_\b \del_\c F \, .
 \eqn\tripled
$$

Let us recapitulate. The  Christoffel connection of an
$n$-dimensional special \K\ manifold splits into a flat holomorphic
piece transforming as a connection and a non-holomorphic term with
tensorial transformation properties. In addition,
one can define another holomorphic connection $A_\a$ which acts
instead on a $(2n+2)$-dimensional symplectic bundle.
This connection is also flat and the special coordinates of special
geometry are the flat coordinates for both connections. It is
important to  distinguish clearly between these two flat holomorphic
connections.

To close this section let us display a few more properties of
$A_\a$ and eq.~\dholo.
Firstly, one easily verifies that $A_\a$ is valued in $sp(2n+2)$:
$QA =(QA)^T$, where $Q$
is the symplectic metric given in \Qsympdef. This is directly related
to the symplectic action \spaction\ on $V$ or similarly to the fact
that $V$ does not contain $(2n+2)$ independent functions. Instead,
$X^A$ and $F_A$ are related by eq.~\hom.
Furthermore,
{}from its definition \cdef\ we learn that $\IC_\a$ satisfies
$$
\IC_\a \IC_\b \IC_\c\IC_\d =0\, , \qquad [\IC_\a, \IC_\b] = 0\, ,
\qquad
\del_{[\a} \IC_{\b]} - A_{[\a} \IC_{\b]} = 0 \, .
\eqn\propc
$$
Thus, $\IC_\a$
generates an Abelian,
$n$-dimensional subalgebra of $sp(2n+2)$ that is nilpotent of order
three. $\IGam_\a$ as defined in eq.~\ac\ also has zero curvature.

Finally, let us observe that eq.~\dholo\ can be turned into a set of
coupled partial differential equations for $V$.
Using eq.~\ac--\cdef\ one finds
$$
{\hat D}_\a {\hat D} _\b(W^{-1})^{{\tilde \c}
\rho\sigma}{\hat D}_{\tilde \c} {\hat D}_\sigma V =0\, ,
\eqn\fp
$$
where ${\tilde \c}$ is not summed over. We see that eq.~\fp\ is
holomorphic and
covariant under \K\ and coordinate transformations. It is this
`solved'
version of \dholo\ which in one dimension turns into the 4th-order
linear differential equation of ref.~\CDGP. Let us turn to this
special case in the following section.

\chap{Ordinary differential equations and $W$--generators on a
one-dimensional special \K\ manifold}
In this section we briefly discuss the  case of  special geometry in
one complex dimension. The reason is that  the
specific example
discussed in ref.~\CDGP\ corresponds to a one-dimensional moduli
space of a  \cy--threefold and so we will be able to make contact
with this example rather easily. Also, in one complex dimension some
further properties of  eq.~\dholo\ will appear.

\def\Dhat{{\hat D}}
In one dimension eq.~\fp\
 reads
$$
\Dhat\Dhat\,W^{-1}\,\Dhat\Dhat V\ =
 \sum_{n=0}^4 a_n(z) \, \del_z^n V\ =\ 0  \ ,
\eqn\oddeq
$$
where $W$ is the
one-dimensional Yukawa coupling. One finds that the coefficients
$a_n$ are not arbitrary but related in the following way
$$
a_3 = 2 \del a_4, \qquad a_4 = W^{-1}, \qquad
a_1 = \del a_2 - \coeff12 \del^2 a_3\, ,
\eqn\acoeff
$$
whereas the coefficients $a_2$ and $a_0$ are complicated functions of
$W$ and the
connections. (Note that in special coordinates eq.~\oddeq\ becomes
very simple and reads
$\del^2\,W^{-1}\,\del^2\,V\ =\ 0$).

The coefficients $a_n$  have to obey well-defined transformation laws
in order to render
eq.~\oddeq\ covariant under coordinate changes ($z\to\tilde z(z)$,
$\del\to\xi^{-1}\del$, $\xi\equiv\del \tilde z/\del z$) as well as
\K\ transformations. It proves convenient to rewrite \oddeq\
slightly. First, one can scale out $a_4$, and furthermore drop the
coefficient proportional to $a_3$ by means of the redefinition
$
V\to \V e^{ - 1/4 \int {a_3(u)\over a_4(u)} du}
$.
This puts the differential equation into the form
$$
(\del^4 + c_2 \del^2 + c_1 \del + c_0)V = 0  \ ,
\eqn\oddeqp
$$
where the new coefficients $c_n$ are combinations of the $a_n$ and
their
derivatives. In this basis $V$ transforms as a $-3/2$ differential,
but the
transformation properties of the $c_n$ are not very illuminating.
However, one
can find combinations of the $c_n$'s and their derivatives which
transform like tensors:
 $$
\eqalign{
\w_2 = &\  c_2  \, , \cr
\w_3 = &\  c_1 -  c_2'\, , \cr
\w_4 = &\   c_0 - \coeff12  c_1'
+ \coeff{1}{5}   c_2''
- \coeff{9}{100}   c_2^2 \ .}
\eqn\wcoeff
$$
A straightforward computation shows
$$
\eqalign{
\tilde\w_2 =&\ \xi^{-2} [ \w_2 - 5\{\tilde z ;z\} ]\, , \cr
\tilde\w_3 = &\ \xi^{-3} \w_3\, , \cr
\tilde\w_4 = &\ \xi^{-4} \w_4 \, ,}
\eqn\wtrans
$$
where $\{\tilde z ; z \}= ( { \del^2 \xi \over \xi} - {3\over2}({\del
\xi\over\xi})^2)$ is the Schwarzian derivative.
($w_2,w_3,w_4$ form a
classical $W_4$--algebra \DIZ, but this
will play no role in the following.) However, eq.~\oddeq\  is not the
most
general 4th-order linear differential equation. Its coefficients
satisfy \acoeff\ and as a consequence one finds $\w_3 = 0$ or
equivalently
$$
\big[\del^4+ w_2 \del^2+w_2'\del+
\coeff3{10}w_2''+\coeff9{100}{w_2}^2+ w_4\big]\,V  = 0 \, .
\eqn\wdeq
$$
Thus, all special geometries in one dimension lead to a  4th-order
linear differential equation that is characterized by  $w_3=0$.
This is due to the fact that the solution vector $V$
does not consist of four completely independent elements, but rather
is restricted by eqs.~\sive\ and \hom.

{}From eq.~\wtrans\ we learn
that there is always a coordinate system in which $w_2=0$ holds. On
the other
hand, $w_3$ and $w_4$ do characterize a  4th-order differential
equation
in any coordinate frame. Thus one can  discuss the properties
of \wdeq\ when in addition
$
w_4\ =\ 0\
$ holds. One finds that this corresponds to
$$
F=  \coeff16{(X^1(z))^3 \over X^0(z)} + c_{AB} X^A X^B \, ,\qquad
\hat D\,W=0\, ,
\eqn\FXrel
$$
where $c_{AB}$ are arbitrary constants. For $c_{AB} =0$, \FXrel\ is
the
$F$--function corresponding to the homogeneous  space
$SU(1,1)/U(1)$ (which
satisfies the stronger constraint $D\,W=0$) \dfnt. Thus, for covariantly
constant Yukawa couplings
the differential equation is essentially reduced to the
differential equation of a torus. This is similar to the situation
for the
$K_3$ surface where the only non--trivial $W$-generator is $w_2$
\LSW.
The possibility of having non--trivial Yukawa couplings, or $w_4 \neq
0$, is
the new ingredient in special geometry. It reflects the possibility
of
having instanton corrections to $W$ or in other words  $w_4$ measures
the deviation from a constant $W$,
which is the
large--radius limit of the \cy\ moduli space.

The significance of the $w$-generators can also be understood in
terms of the first order equation \dholo. Any linear 4th-order
differential equation can be cast into the form \dholo\ with
$$A\ =\  \pmatrix{
0  & 1 &  0 & 0 \cr
-\coeff3{10}w_2 & 0  & 1 & 0 \cr
-\shalf w_3 & -\coeff4{10}w_2 & 0 & 1\cr
 -w_4 &-\shalf w_3 &-\coeff3{10}w_2 & 0\cr}\ \in sl(4,R)\ .\eqn\wmat
$$
To understand this form, recall the well-known relationship\foot{We
thank R.Stora for discussions on this point.} between $W$-algebras
and a
special, ``principally embedded'' $SL(2)$ subgroup $\cS$ \Kos\ of
$G=SL(N)$ (in
fact, $G$ can be any simple Lie group). The generators of $\cS$ are
$$
J_-=\sum_{{simple\atop roots\ \a}} b_\a E_\a\ ,\ \ \
J_+=\sum_{{ simple\atop roots\ \a}} c_\a(b_\a) E_{-\a}\ ,\ \ \
J_0=\rho_G\cdot H\ ,\eqn\gendef
$$
where $b_\a$ are arbitrary non-zero constants, $c_\a$ depend on
the $b_\a$ in a certain way and $\rho_G$ is the Weyl vector.
An intriguing property \Kos\ of $\cS$ is
that the adjoint of any group $G$ decomposes under $\cS$ in a
very specific manner:
$$
adj(G)\to \bigoplus r_j\ ,\eqn\sdecomp
$$
where $r_j$ are representations of $SL(2)$ labelled by spin $j$,
and the values of $j$ that appear on the r.h.s. are
equal to the exponents of $G$. The exponents are just the degrees
of the independent Casimirs of $G$ minus one (for $SL(N)$, they
are equal to $1,2,\dots,N-1$).

Recalling that the Casimirs are one-to-one
to the $W$ generators associated with $G$, one easily sees that
the decomposition \sdecomp\ corresponds to
writing the connection \wmat\ in terms of $W$-generators;
more precisely, for an $N$th-order equation
related to $G=SL(N)$, the connection \wmat\ can be written as
\doubref\BFOFW\DIZ:
$$
A\ =\ J_--\sum_{m=1}^{N-1} w_{m+1} (J_+)^m\ ,\eqn\wdec
$$
where $J_\pm$ are the $SL(2)$ step generators \gendef\ (up to
irrelevant
normalization of the $w_n$).

In our case\foot {The choice \wmat\ for $A$ corresponds to an
embedding
\gendef\ with $b_1=b_2=b_3=1, $ and $c_1=c_3=3/10, c_2=4/10$.} with
$N=4$, the
decomposition \sdecomp\ of the adjoint of $SL(4)$ is given by
$j=1,2,3$, which
corresponds to $w_2,w_3$ and $w_4$. We noticed above that
$w_3\equiv0$ for
special geometry and this means that $A$ belongs to a Lie
algebra that
decomposes as $j=1,2$ under $\cS$. It follows that this Lie algebra
is $sp(4)$.
Indeed, remembering that the algebra $sp(n)$ is spanned by matrices
$A$ that
satisfy $A\,Q + Q\,A^T=0$, we can immediately see from \wmat\ that
$$
A\ \in\ sp(4)\ \qquad \ \ \longleftrightarrow\ \ \qquad \
  w_3\ \equiv\ 0\  .\eqn\spexpl
$$
Above, the symplectic metric $Q$ is taken as in \Qsympdef.

Similarly, if in addition  $w_4=0$ (which corresponds to a
covariantly
constant Yukawa coupling),
$A$ further reduces to an $SL(2)$ connection.
This $SL(2)$ is identical to the principal $SL(2)$ subgroup, $\cS$,
since according to \wdec\ the entries labelled by $w_2$ and $1$ in
\wmat\ are directly given by the $\cS$ generators $J_+$ and $J_-$.

\chap{Relation with the moduli space of \cy\ threefolds}
It is well known that the moduli space of \cy\ threefolds $\cM$ is
 a special \K\  manifold \refs{\FSa ,\strominger,\CD} and thus
our considerations of the previous sections
immediately apply.
In particular,
${\bf U} = (V,U_\a,\bar U_{\bar\b},\bar V)$  can be identified with
the basis elements of the
third (real) cohomology of $\cM$,
$H^3= H^{(3,0)} \oplus H^{(2,1)} \oplus H^{(1,2)} \oplus H^{(0,3)}$
\doubref\strominger\CD. Furthermore, $V = (X^A, F_A)$  are just the
periods of
the holomorphic three--form $\Omega$
\refs{\Griffiths{,}\Candelas{,}\CD}:
$$
X^A\ =\ \int_{\gamma_A}\Omega\ ,\ \qquad F_B\ =\
\int_{\gamma_B}\Omega\, , \qquad A,B=0,\dots,n\, .
\eqn\Omperiods
$$
Here, $\c_A, \c_B$ are the usual basis cycles of $H_3$. Consequently,
${\bf U}$  corresponds to the period matrix of
$\cM$.

The period matrix is defined only
up
to local gauge transformations,
which are precisely of the form \gaugetrans\ \Griffiths. Thus,
from the considerations of section~3 it immediately follows
that
the period matrix can also be presented in the holomorphic gauge
\puregauge.\foot{
In addition, the period matrix is equivalent under conjugation by an
integral
matrix, $\Lambda$: ${\bf U}\sim{\bf U}\Lambda$. These transformations
$\Lambda\in
Sp(2n+2,\ZZ)$, which correspond to changes of integral homology
bases, preserve
the symplectic bilinear intersection form $Q$ of $H_3(\cM,\ZZ)$, that
is:
$\Lambda\,Q\,\Lambda^T=Q$. The subset of these transformations that
leave $F$
invariant up to redefinitions constitute the ``duality group''.}
Eq.~\dholo\ exactly corresponds to the \pf\ equations obeyed by the
periods \refs{\CF{--}\morrison}.
An explicit form of $\Omega$ can be obtained in terms of the defining
polynomial of the \cy\ manifold \refs{\Candelas,\tpcandelas}.

Ref.~\CDGP\ considered a particular
\cy\ manifold (a quintic in $CP_4$) whose moduli space is
one-dimensional.\foot{Subsequently, this computation was generalized
for a few other examples of \cy\ manifolds with one-dimensional
moduli spaces in refs.~\refs{\morrison,\FKT}.}
$X^A$ and $F_A$ were obtained by
explicitly evaluating the period integrals \Omperiods. It was then
noted that the periods do satisfy a 4th-order
holomorphic differential equation. This differential equation is
exactly eq.~\oddeq\ with specific coefficients $a_n$ which indeed
satisfy \acoeff.
However, for moduli spaces of arbitrary dimension it might be easier
to solve the differential equation rather than performing the
integrals \Omperiods. Therefore, let us now turn to an alternative
method of computing $V$.

\chap{Computation of $K$ and $\WABC$}
So far we  worked  within the context of special
geometry which is the framework for the \el\ of $(2,2)$ vacua.
In this section we indicate how to explicitly compute the
\K\ potential and the Yukawa couplings for such a string vacuum.
Various different strategies have been employed. In SCFTs
$C_{\a\b\c}$ is
computed
via a (moduli dependent) three-point function of the ${\bf 27} ({\bf
\bar{27}})$ matter fields \twotworeview.
This can often only be done perturbatively
in the moduli fields around a particular point in moduli space
\Cvetic.
 Once  $C_{\a\b\c}$  as a function of the moduli is known,
eq.~\curvb\ can be used as a differential
equation which
determines $K$. This strategy has been proposed in ref.~\DKLa\ and in
sufficiently simple examples $K$ can indeed be calculated. (In TCFT
the
same strategy has also been used to determine the metric of
``topological--antitopological fusion'' \CV. We will come back to
this point in the next section.)

As we already mentioned in the last section it is sometimes  possible
to explicitly evaluate eqs.~\Omperiods. In this section we outline
an alternative procedure  for computing $V$
which was advocated in refs.~\refs{\CF,\LSW}. The idea is to
calculate
$A_\a$ of eq.~\dholo\ and then solve the resulting differential
equation for $V$ which determines $K$ and
$C_{\a\b\c}$ via eqs.~\kdef\ and \fabc.
This computation  is particularly simple
for the  class of $N=2$ string vacua which can be represented by a
\LG\ superpotential $\W$. Therefore, let us recall a few basic
properties of $\W$. (For a more exhaustive
review see for example \refs{\ntworeview}.)
The unperturbed superpotential $\W_0$ is a quasi-homogeneous function
of the
chiral superfields $x^i,i=1,\ldots,N$. The  chiral primary operators
of the SCFT are represented by all monomials of $x^i$ modulo the
equation of motion $\del_{x_i} \W_0=0$. These operators form a so
called chiral ring where
the ring multiplication is identified with polynomial multiplication.
Maintaining conformal invariance $\W_0$ can be perturbed by exactly
marginal operators $p_\a(x_i)$. The perturbed superpotential $\W$
then reads
$$
\W(x_i,z^\a)\ =\ \W_0(x_i) - \sum z^\a\,p_\a(x_i)\ ,\qquad\ \
\a=1,\dots,n\ ,
\eqn\suppot
$$
where $z^\a$ are
the (dimensionless) moduli parameters. In a more general situation
one can add a perturbation of a relevant operator which induces a RG
flow
to another SCFT. Here we only focus on  marginal perturbations since
they correspond to massless fields in the \leel. The marginal
operators among
themselves generate a chiral $(2n+2)$ dimensional (sub)-ring
$\{1,p_\a,p^\b,\rho \}$
where $\rho$ is the unique
top element of the chiral ring and
$p^\b$ is defined by $p_\a\, p^\b=\delta^\b_\a\rho$. In this
perturbed ring one defines polynomial multiplication  via
$$
p_\a \ p_\b=W^{(p)}_{\a\b\c}(z)\ p^\c\quad {\rm mod} \, \del_{x_i}
\W\, .
\eqn\polyp
$$
The non-trivial point is that $V$ (defined in eq.~\sive) can be
expressed in terms of $\W$ as follows
\refs{\Candelas,\CF,\LSW,\tpcandelas}:
$$
V =\ \int_{\gamma_{A},\gamma_B}\,{ 1\over
 \W^{\ell}(x_i,z^\a)}\, \omega\ , \qquad
 \omega = \sum_i^N (-1)^i x^i dx^1\ldots \hat{dx}^i\ldots dx^N \, ,
\eqn\resid
$$
where $\ell = (N-3)/2$.\foot{I like to thank P.~Candelas and
W.~Lerche
for discussions on this point.}
(The precise definition of the integral is given in
\refs{\Candelas,\LSW}, it is not important in the following.)
\def\per#1.#2{\int\!\coe#1.\W^{\ell+#2}.\, \omega}
Then
the $(2n+2)\times (2n+2)$ dimensional matrix ${\bf V}$ (defined in
eq.~\puregauge) can  be represented as
$$
{\bf V}\ =\ \pmatrix{\per1.1\cr\per p_\b.2\cr\per
p^\c.3\cr\per\rho.4\cr}\ .
\eqn\vholo
$$
Using eqs.~\polyp\ and \vholo\
one easily verifies that ${\bf V}$ indeed satisfies eq.~\dholo. For a
given $\W$ we can use the representation \vholo\ and explicitly
compute $A_\a$ as a function of $z^\a$ by taking derivatives of ${\bf
V}$ and rewriting it as $A_\a \cdot {\bf V}$. Then eq.~\dholo\ turns
into a
non-trivial differential equation for $V$ which (at least in
principle) can be solved. For the quintic of
ref.~\CDGP\ $A_\a$ has been computed in
refs.~\refs{\CF,\LSW} and the solution of  eq.~\oddeq\ for this
example is discussed in \refs{\CDGP,\tpcandelas} and so we refrain
here from repeating this analysis.\foot{Further examples are
discussed in refs.~\refs{\morrison,\FKT}.} The important point  we
want to stress  is
that $A_\a$ is determined  from $\W$ alone. Thus it is not always
necessary to solve the SCFT in order to  determine the tree-level
couplings of the \el.
Rather, the necessary information about the couplings is already
encoded in the \LG\ superpotential $\W$.

In general $A_\a$ will not come out in the form \ac. However,
as we noted in the last section eq.~\dholo\ still displays a gauge
covariance with a holomorphic matrix $S$ of the form \gaugetrans.
This gauge freedom  can be used to put $A_\a$ into the form
$A_\a=\IGam_\a +\IC_a$ where $\IGam_\a$ and $\IC_a$ are given by
\gammahat\ and \cdef. The matrices $\IC_\a$ (which contain
$W_{\a\b\c}$)
can be viewed as the structure constants of the $(2n+2)$-dimensional
ring generated by $\{1,p_\a,p^\b,\rho \}$.
By going to the gauge \cdef\ one finds a relationship between the
Yukawa couplings $W_{\a\b\c}$ and $W^{(p)}_{\a\b\c}$ of eq.~\polyp.
However, this relation is still not unique.
It is clear from eq.~\ktransxf\ that the Yukawa couplings are defined
only up
to a \K\ transformation $W_{\a\b\c}\rightarrow W_{\a\b\c} e^{-2f}$
and we already remarked that eq.~\dholo\ is covariant under \K\
transformations. Of course, any physical coupling is (\K-) gauge
independent. Thus, one has to determine $\WABC$ and $K$
in the same \K\ gauge and that is exactly what eq.~\dholo\ does.
The solution of \dholo\ determines $K$ in the same gauge we have
chosen for $\WABC$.

A slightly different strategy is to solve $\IGam_\a =0$, which
determines the flat
coordinates,
 instead of solving the differential
equation for $V$.  As was shown in
detail in \LSW, imposing this condition gives a non-linear
differential equation
that
determines explicitly the dependence of the \LG couplings $z^\a$ on
the
$t^\a$. (This is closely related to the
mirror map of ref.~\CDGP.)
 Once we have
$W_{\a\b\c}$ in flat coordinates,   $F$ is determined via
eq.~\tripled\ up to three integration constants.
These correspond to the initial conditions of eq.~\dholo.
Unfortunately, they contribute to the physical Yukawa
couplings
via $e^K$. In ref.~\CDGP\ they were determined by using the mirror
hypothesis \mirror\ and the knowledge of $F$ in the large radius
limit. In this limit they can be interpreted as
perturbative $\sigma$--model loop corrections in Calabi--Yau
compactifications \CDGP.

Finally,
we should mention a disadvantage of the present approach. Not all
moduli of a given SCFT can be represented in such a simple fashion as
in eq.~\suppot. Some of them  appear in so-called twisted sectors,
and one has to  use mirror symmetry to get further information about
these twisted moduli. However, the method outlined here allows for
the determination of $K$ and
$C_{\a\b\c}$ of a larger class of string vacua and a larger part of
the moduli space than previously known.

\chap{Conclusions}

Special \K\ geometry also made its appearance in topological
conformal field theories (TCFT)
\refs{\DVV,\CV, \top} and in this final section we briefly  comment
on this aspect.

Every $N=2$  SCFT can be `twisted' into a TCFT \tntwo\ which leads to
a projection onto the chiral primary
fields $\phi_i$ as the only physical operators of the TCFT. A family
of TCFT is defined by the action
$
S = S_0 + \sum_i t^i \int \phi_i
$, where  $t^i$ are the
corresponding (complex) coupling parameters. All correlation
functions in the TCFT can be expressed in terms of $\eta_{ij} \equiv
\vev{
\phi_i\phi_j}$ and $W^{\rm top}_{ijk} \equiv \vev{
\phi_i\phi_j\phi_k}$.
 By using the Ward identities of the TCFT one   finds \DVV
 $$
\del_k \eta_{ij} =0 \, , \qquad  W^{\rm top}_{ijk} =
\del_i\del_j\del_k F^{\rm top} \, ,
\eqn\toppro
$$
which are precisely
 the properties of special geometry in flat coordinates
(eqs.~\flutstuff, \tripled). However, the action has been perturbed
by
 all chiral primaries including
the relevant operators of the theory.  The moduli space of
$(2,2)$ SCFT which we discussed so far corresponds  to the subspace
of the marginal
deformations of the TCFT and the (holomorphic) Yukawa couplings
coincide (up to the \K\ gauge freedom discussed above) with the
topological correlators $ W^{\rm top}_{ijk}$ \refs{\GS,\CDFLL}.

For the  minimal models  $W^{\rm top}_{ijk}$ has been determined
using a \LG\ representation exactly analogous to eq.~\suppot, where
$\a$ now runs over all topological deformations.
The flat coordinates $t^a$   arise as the solution of a Lax equation
of the generalized KdV hierarchy where the \LG\ superpotential $\W$
is identified with the Lax operator and $F$ plays the role of the
$\tau$-function.
This corresponds to $\IGam_\a = 0$ where $\IGam_\a$ in this context
is the Gauss--Manin connection \refs{\LSW,\top}.
It would be interesting to see what the analogous statement for a
TCFT corresponding to a string vacuum is, for example for the quintic
of ref.~\CDGP. A step in this direction has been reported here at
this workshop by B.~Dubrovin \dubrovin\ where the integrability of
special geometry is shown.

A slightly different perspective was pursued in ref.~\CV.
  It was shown  that the analogue of the \K\ metric
$g_{i\bar j}$ arises in the ``fusion'' of a TCFT with its
anti-topological ``partner''. This metric  $g_{i\bar j}$ also
satisfies eq.~\curvb\ and \dc\ and thus is a metric on a
(generalized) special manifold.
Exactly  as before the structure of special geometry is being
extended to the much
bigger space of all topological deformations. However, when
 relevant perturbations are present the metric $g_{i\bar j}$ can no
longer be expressed in terms of  holomorphic objects as is the case
on the subspace of marginal perturbations.

The discovery of special geometry  in TCFT were partly the motivation
for the investigation of refs.~\refs{\FL,\CDFLL}. We hope to have
clarified the structure of the subspace of marginal perturbations
which is the subspace relevant for the \el\ of $(2,2)$ string vacua.
We should also
mention that not only the \pf\ equations arise from these topological
considerations, but it seems that there are further properties of the
\leel\ encoded in
some appropriate topological field theory \AMW. Clearly, this
deserves further study.

Finally, we did not touch upon  the quantum duality
symmetry which imposes a strong constraint on the couplings in the
\el. Clearly, the duality group is closely related to the
monodromy group of the differential equation \LSW,  which in turn
depends on the zeros and poles of
the Yukawa couplings. It would be worthwhile to make the relation
between the duality group and the monodromy group more precise.

\ack
I would like to thank A.~Ceresole, R.~D'Auria, S.~Ferrara and
W.~Lerche for a very enjoyable collaboration. I also greatly
benefited from discussions with P.~Berglund, P.~Candelas, X.~de la
Ossa, B.\ de Wit, B.~Dubrovin, A.\ Klemm, R.~Schimmrigk and  S.\
Theisen.
I thank the organizers of the workshop for providing such a
stimulating environment.

\refout

\end